\documentclass[11pt]{article} 
\usepackage[utf8]{inputenc} 

\usepackage{geometry} 
\usepackage{graphicx}

\usepackage{booktabs}
\usepackage{array} 
\usepackage{paralist} 
\usepackage{verbatim} 
\usepackage{subfig} 

\usepackage{fancyhdr}
\usepackage{sectsty}
\allsectionsfont{\sffamily\mdseries\upshape} 

\usepackage[nottoc,notlof,notlot]{tocbibind} 
\usepackage[titles,subfigure]{tocloft}

\title{Geometrisation of Fermions in Semi-Riemannian Spacetimes}
\author{William J. Leigh\thanks{wl197@leicester.ac.uk} \\Division of Library and Learning Services\\University of Leicester, University Road, Leicester LE1 7RH\\United Kingdom}
\date{April 2024}

\begin{document}
\maketitle

\begin{abstract}An earlier scheme [arXiv:2404.03360], where torsion plays an essential part in a flat spacetime account of fermion spin, is extended to spacetimes with non-zero Riemann curvature. It is found that further essential features of the fermion, in particular its electromagnetic field, are determined by the curvature and torsion of spacetime. A natural model for the metric, singular at the spin axis of the particle, enables integration of a stress-energy tensor that fixes a realistic electron mass in terms of the Planck mass, despite the many orders of magnitude by which these values differ. Although fields here are not explicitly quantised, it is hoped that the normally overlooked, yet underlying and clearly fundamental geometric structures, may account for the apparent incompleteness often believed to characterise quantum theories, as evidenced in the well-known Einsten, Podolsky, Rosen paradox.\end{abstract}


\section{Introduction}\label{Ch1}

The analysis of the semi-Riemannian spacetime and its fields presented in this paper is carried out in explicit recognition of a proposed generalised conformality between it and the author's earlier flat spacetime approach \cite{leigh}: not only will spacetime metrics be conformal {\em sensu stricto}, but the Riemann analysis will be guided by the attempt to express all its fields as real multiples of their flat, Cochlean spacetime\footnote{See \cite{leigh}, which will be referred to in the text as Paper 1, for the definition of Cochlean spacetime.} counterparts. 

Section \ref{Ch2} begins by proposing a Riemann physical spinor resembling the Cochlean version. In contrast with the flat-space case however, the differential identity this spinor is taken to satisfy is essentially trivial, expressing as it does a hypothesised Riemann constancy of the spinor field. The vanishing of the spinor derivative then determines the general form of the Riemann spinor connection, in which a real vector field $A_\mu$ appears in a way similar to that in which its putative Cochlean counterpart $a_\mu$, the electromagnetic potential, appeared in Paper 1. The Riemann spinorial formulation is completed by examining the conformal factors of the spin metric and Infeld van der Waerden symbols [IvdW], before considering how the current associated with the physical spinor might be a multiple of the Cochlean current; it is found that a transformation of the Riemann IvdW is required that is essentially the same as that used in Paper 1.

Section \ref{Ch3}  examines the Riemann tensor connection that is determined by requiring covariant constancy of the IvdW, a connection displaying a torsion that turns out to extend that of the Cochlean geometry. The Riemann torsion however, when taken together with the pseudoequality of $A_\mu$ following from its being supposed a multiple of the Cochlean $a_\mu$, ensures that setting $A_4$ to be a constant multiple of the metric conformal factor causes the Riemann electromagnetic field to vanish. In this way the electromagnetic field of a particle becomes an epistemologically Poincaré phenomenon that depends upon a spacetime curvature and torsion of which a Poincaré observer\footnote{A ‘Poincaré observer’ is understood here to be a physicist who insists that spacetime has the standard isotropic and homogeneous Minkowski form, while a ‘Newtonian observer’ insists that spacetime is simply a direct product of 3-dimensional Euclidean space with a 1-dimensional homogeneous time.} remains unaware, much as the gravitational field of a body is an epistemologically Newtonian phenomenon that depends upon a spacetime curvature of which a Newtonian observer equally remains unaware.

Section \ref{Ch4}  defines a Riemann energy-momentum tensor, $T_{\mu\nu}$, based upon an Einstein tensor constructed from the symmetrised Ricci tensor. Taking the physical interpretation of the Cochlean $t_{\mu\nu}$ to inform the structure of $T_{\mu\nu}$ suggests that certain conditions be imposed on the latter, and these constrain not only otherwise arbitrary Riemann fields originally introduced in the definition of the spinor connection, but also the metric conformal factor, which turns out to depend only upon the spatial cylinder radial coordinate.Taking all constraints into account, Section \ref{Ch4} ends with a form of the Riemann energy density tensor that incorporates arbitrary functions of the spacetime variables $\rho$ and $z$ which can be chosen to reproduce its Cochlean counterpart. Whilst this demonstrates the mutual consistency of flat and curved spacetime formulations, the arbitrariness it contains leaves open the question of a mass scale. 

Section \ref{Ch5} uses a simple model for the metric conformal factor that incorporates the natural Planck mass scale of gravitation (when combined with electromagnetism and quantum mechanics) into a self-consistency condition that determines the mass of an idealised fermion, thereby completing the purely formulational approach of the current paper. 

\section{The Riemann Spinor Formulation}\label{Ch2}

\subsection{The Physical Spinor}

The principal requirement of the Riemann physical spinor is that it describe a stationary, stable, spin-1/2 fermion at the origin of the spatial coordinate system; as will appear in the current and following sections, such a spinor can be used to determine a large part of the geometry of spacetime. Given the azimuthal symmetry, and following the Cochlean pattern, it is reasonable to propose the physical spinor form
\begin{equation}\label{(2.1.1)}{\Xi _A} = {M^{3/2}}S\left( {\begin{array}{*{20}{c}}
{{e^{i\Phi }}}\\
{{e^{ - i\Phi }}}
\end{array}} \right),\;\quad \Phi  := \phi /2 - Mt,\end{equation}
where $M$ is a parameter with the dimension of mass, and the modulus, $S$, of the spinor components is a function only of the cylinder space variables $\rho$ and $z$. 
It may be recalled that the time-dependence of the Cochlean spinor was modelled on the standard quantum field theoretic oscillation associated with mass $m$, and a similar parameter appears in (\ref{(2.1.1)}) where it further offers the convenience of using a dimensionless $S$. Physical interpretation of this parameter will however be deferred until the impact can be assessed of the natural mass scale of General Relativity when it is used with quantum theory,  based as it then becomes upon an Einstein condition that incorporates the Planck mass. 

\subsection{The Spinor Connection}

In the absence at the strictly formulational stage of a mass scale, it will be insisted that the Riemann physical spinor have vanishing covariant derivative, 
\begin{equation}\label{(2.2.1)}{D_\mu }{\Xi _A} = 0 \Leftrightarrow {d_\mu }{\Xi _A} - \Gamma _{\mu A}^{\quad  \bullet B}{\Xi _B} = 0,\end{equation}
when it can be shown that the spinor connection may be written as the sum 
\begin{equation}\label{(2.2.2)}\Gamma _{\mu A}^{\quad  \bullet B} = \left[ {{{\left( {\ln S} \right)}_\mu } - ie{A_\mu }} \right]E_A^{\; \bullet B} + \left( {\begin{array}{*{20}{c}}
{i{\Phi _\mu } + {U_\mu }}&{{e^{2i\chi }}\left( {ie{A_\mu } - {U_\mu }} \right)}\\
{{e^{ - 2i\chi }}\left( {ie{A_\mu } + {U_\mu }} \right)}&{ - i{\Phi _\mu } - {U_\mu }}
\end{array}} \right),\end{equation}
where $U_\mu$ and $A_\mu$ are arbitrary complex fields. In order to generate a steady, axially symmetric Cochlean electromagnetic field, the $A_\mu$ here [which will have the Cochlean electromagnetic potential $a_\mu$ as its counterpart] must be real and depend only upon $\rho$ and $z$; a similar coordinate dependence will be assumed for the complex $U_\mu$. When the spin metric $E_{AB}$, treated in the next section, is used to lower the index B in (\ref{(2.2.2)}) by postmultiplication, the principal addends on the RHS become respectively skew and symmetric in their spinor indices.

\subsection{The Spin Metric}

As just noted, this fundamental element of the Riemann spinor geometry in its lower index form, $E_{AB}$ , is taken to lower spin indices by multiplication on the right and, since it is skew, is necessarily a multiple of the Poincaré form $e_{AB}$, for which the convention of Bade and Jehle \cite{badejehle} that ${e_{AB}} = \left( {\begin{array}{*{20}{c}}
0&1\\
{ - 1}&0
\end{array}} \right)$  is followed here, so the Riemann form becomes
\[{E_{AB}} = \mathop f\limits^s {e_{AB}} = \mathop F\limits^s {e^{i\alpha }}{e_{AB}};\quad \mathop F\limits^s  := \left| {\mathop f\limits^s } \right|,\;\alpha  := \arg \left( {\mathop f\limits^s } \right).\]
For the same reasons as in the Cochlean case, repeated below, it is necessary to insist on covariant constancy of the product of the metric and its conjugate, 
\begin{equation}\label{(2.3.2)}\begin{array}{c}
0 = {D_\mu }\left( {{E_{AB}}{E_{\dot C\dot D}}} \right) = {d_\mu }\left( {{{\left| {\mathop f\limits^s } \right|}^2}{e_{AB}}{e_{\dot C\dot D}}} \right) - 2{\Gamma _{\mu [AB]}}{E_{\dot C\dot D}} - 2{E_{AB}}{\Gamma _{\mu [\dot C\dot D]}}\\
 = 2\left( {{{\left( {\ln \left| {\mathop f\limits^s } \right|} \right)}_\mu } - 2{{\left( {\ln S} \right)}_\mu }} \right){E_{AB}}{E_{\dot C\dot D}},
\end{array}\end{equation}
so that
\begin{equation}\label{(2.3.3)}{E_{AB}} = {S^2}{e_{AB}},\end{equation}
where, in the absence as yet of a spinor normalisation condition, an overall arbitrary positive constant has been absorbed into $S^2$ and, in line with the drive for generalised conformality, the phase $\alpha$ has been set to zero. It should be noted too that, reflecting perhaps well known difficulties in localising energy densities in Riemann spacetimes, the Riemann spinor modulus $S$ must, in contrast with the Cochlean form $H$, which vanishes asymptotically, become unity in the asymptotic region if flat-space spinor structure is to be recovered. It then follows that $H$ and $S$, although each must be understood as associated with the same spatially confined particle at the origin, cannot be mutually conformal, but may more naturally be related by exponentiation. 

\subsection{Other Conformal Factors}

There are three factors here that can be termed conformal, namely for the Infeld van der Waerden symbols, for the spin metric and – with the traditional meaning of the term – for the metric of the spacetime. They must all be consistent with the condition that defines the IvdW, 
\begin{equation}\label{(2.4.1)}\Sigma _{\mu A}^{\quad  \bullet \dot X}\Sigma _{\dot X}^{\nu  \bullet B} + \Sigma _A^{\nu  \bullet \dot X}\Sigma _{\mu \dot X}^{\quad  \bullet B} =  - G_\mu ^\nu E_A^{\;B} =  - \delta _\mu ^\nu \delta _A^{\;B},\;{G_{\mu \nu }} = {\Omega ^2}{g_{\mu \nu }} = {e^{2\omega }}{g_{\mu \nu }}.\end{equation}
As already exemplified, in this paper upper and lower case forms of the same symbol are generally either the Riemann and Cochlean versions respectively of the same thing, as here with $G$ and $g$, or they are related by exponentiation, as with $\Omega$ and $\omega$; which is intended is determined by context. The Cochlean metric is taken to be the standard Minkowski cylinder polar form, ${g_{\mu \nu }} = \left( { - 1, - {\rho ^2}, - 1,1} \right)$, and $\Omega$ is assumed real and positive [see  (\ref{(2.4.4)}) below].
Lowering the spacetime index and raising and lowering the dotted spin index gives
\[{\Sigma _{\mu A\dot X}}\Sigma _\nu ^{\dot XB} + {\Sigma _{\nu A\dot X}}\Sigma _\mu ^{\dot XB} = {G_{\mu \nu }}E_A^{\;B},\]
and from this form it is clear why, in addition to the covariant constancy of an IvdW with, say, oppositely placed spacetime and spin indices, it is essential that the product of the spin metric and its conjugate be covariantly constant, as in (\ref{(2.3.2)}), since it ensures the covariant constancy of the spacetime metric, or equivalently, the vanishing of the tensor of non-metricity\footnote{For this formulation and the general notational background for this section see \cite{badejehle}, \cite{penrose02}, \cite{poplawski}.}.

The relationship between Riemann and Poincaré/Cochlean IvdW will be fixed here by first introducing a complex mixed constant factor, as in 
\begin{equation}\label{(2.4.3)}\Sigma _{\mu A}^{\quad  \bullet \dot X} = \mathop f\limits^m \sigma _{\mu A}^{\quad  \bullet \dot X},\end{equation}
when, with the conformal factor of the tensor metric defined in  (\ref{(2.4.1)}), there follows
\begin{equation}\label{(2.4.4)}{\left| {\mathop f\limits^m } \right|^2}{\Omega ^{ - 2}}\left[ {\sigma _{\mu A}^{\quad  \bullet \dot X}\sigma _{\dot X}^{\nu  \bullet B} + \sigma _A^{\nu  \bullet \dot X}\sigma _{\mu \dot X}^{\quad  \bullet B}} \right] =  - \delta _\mu ^{\;\nu }\delta _A^{\;B} \Rightarrow \left| {\mathop f\limits^m } \right| = \Omega. \end{equation}
IvdW with spin indices in the same position must be hermitian in order to preserve the standard correspondence between hermitian second order spinors and real four vectors, and it then follows by either raising or lowering a spin index in  (\ref{(2.4.3)}) that $\mathop f\limits^m $  should be chosen real, and $\mathop f\limits^m  = \Omega $ will therefore be taken.
It now follows that for the Riemann forms of the IvdW with oppositely placed spin and spacetime indices
\[\Sigma _\mu ^{A\dot X} = \mathop f\limits^m {\left( {\mathop f\limits^s } \right)^{ - 1}}\sigma _\mu ^{A\dot X} = \Omega {S^{ - 2}}\sigma _\mu ^{A\dot X},\quad \Sigma _{A\dot X}^\mu  = {\Omega ^{ - 1}}{S^2}\sigma _{A\dot X}^\mu, \]
whilst for similarly placed indices
\[{\Sigma _{\mu A\dot X}} = \Omega {S^2}{\sigma _{\mu A\dot X}},\quad \;{\Sigma ^{\mu A\dot X}} = {\Omega ^{ - 1}}{S^{ - 2}}{\sigma ^{\mu A\dot X}}. \]

\subsection{The Physical Spinor Current}

The obvious candidate for the Riemann current associated with the physical spinor is the real four-vector, sesquilinear in the spinor, given by
\begin{equation}\label{(2.5.1)}{J_\mu } := \Sigma _\mu ^{A\dot B}{\Xi _A}{\Xi _{\dot B}},\end{equation}
and it is this density that imposes the dimensional factor $M^{3/2}$ in  (\ref{(2.1.1)}).
The current proposed here will be Riemann constant, since the IvdW1 will be taken constant to determine the tensor connection, whilst  (\ref{(2.2.1)}) is a fundamental assumption for the physical spinor. It is important to consider whether a Cochlean current could be a conformal multiple of  (\ref{(2.5.1)}). A reasonable Cochlean understanding of the physical situation will require not only asymptotic vanishing of its current, but also that that current have non-vanishing time and azimuthal components only, whilst these components must be independent of both time and azimuth; these conditions also ensure that the Cochlean divergence of the current vanishes. 
It is not difficult to see that this will be impossible if the polar forms of the Riemann IvdW are simply conformal to the standard Poincaré versions, which are
\[\begin{array}{l}
\sigma _{1A}^{\quad  \bullet \dot B} = \sigma _{\rho A}^{\quad  \bullet \dot B} = \frac{1}{{\sqrt 2 }}\left( {\begin{array}{*{20}{c}}
{{e^{i\phi }}}&0\\
0&{ - {e^{ - i\phi }}}
\end{array}} \right);\;\sigma _{3A}^{\quad  \bullet \dot B} = \sigma _{zA}^{\quad  \bullet \dot B} = \frac{{ - 1}}{{\sqrt 2 }}\left( {\begin{array}{*{20}{c}}
0&1\\
1&0
\end{array}} \right);\\ \\ 
\sigma _{2A}^{\quad  \bullet \dot B} = \sigma _{\phi A}^{\quad  \bullet \dot B} = \frac{1}{{\sqrt 2 }}\left( {\begin{array}{*{20}{c}}
{i\rho {e^{i\phi }}}&0\\
0&{i\rho {e^{i\phi }}}
\end{array}} \right);\;\sigma _{4A}^{\quad  \bullet \dot B} = \sigma _{tA}^{\quad  \bullet \dot B} = \frac{1}{{\sqrt 2 }}\left( {\begin{array}{*{20}{c}}
0&{ - 1}\\
1&0
\end{array}} \right).
\end{array} \]
To demonstrate this, and to establish a modification of the IvdW that resolves this problem, consider the transformation, effectively already used in Paper 1, defined by
\[\;\hat \sigma _{\mu A}^{\quad  \bullet \dot B} := T\sigma _{\mu A}^{\quad  \bullet \dot B}T,\quad T := \left( {\begin{array}{*{20}{c}}
{\exp i\left( {\theta  - \mu t} \right)}&0\\
0&{\exp  - i\left( {\theta  - \mu t} \right)}
\end{array}} \right),\]
where $\theta$ and $\mu$ are real.
The transformed IvdW1 satisfy the Cochlean/Poincaré version of the purely algebraic defining condition  (\ref{(2.4.1)}), and this remains true for the Riemann versions, obtained by applying the conformal factor in  (\ref{(2.4.3)}).
Omitting a real factor of $ - \frac{{{M^3}}}{{\sqrt 2 }}\Omega {S^{ - 2}}$, the transformed IvdW give the Riemann current components to be, using here the definition  $\zeta  := S\exp i\Theta $, 
\begin{eqnarray}
{J_\rho }& \sim& \left( {\begin{array}{*{20}{c}}
{{\zeta ^ * }{e^{ - i\Phi }}}&{ - \zeta {e^{i\Phi }}}
\end{array}} \right)\left( {\begin{array}{*{20}{c}}
{{e^{2i\left[ {\Phi  + \theta  + \left( {M - \mu } \right)t} \right]}}}&0\\
0&{ - {e^{ - 2i\left[ {\Phi  + \theta  + \left( {M - \mu } \right)t} \right]}}}
\end{array}} \right)\left( {\begin{array}{*{20}{c}}
{{\zeta ^ * }{e^{ - i\Phi }}}\\
{\zeta {e^{i\Phi }}}
\end{array}} \right)\nonumber\\
& =&  2{S^2}\cos 2\left[ {\Theta  - \theta  - \left( {M - \mu } \right)t} \right] \nonumber
\end{eqnarray}
\begin{eqnarray}
{J_\phi }&\sim&i\rho \left( {\begin{array}{*{20}{c}}
{{\zeta ^ * }{e^{ - i\Phi }}}&{ - \zeta {e^{i\Phi }}}
\end{array}} \right)\left( {\begin{array}{*{20}{c}}
{{e^{2i\left[ {\Phi  + \theta  + \left( {M - \mu } \right)t} \right]}}}&0\\
0&{{e^{ - 2i\left[ {\Phi  + \theta  + \left( {M - \mu } \right)t} \right]}}}
\end{array}} \right)\left( {\begin{array}{*{20}{c}}
{{\zeta ^ * }{e^{ - i\Phi }}}\\
{\zeta {e^{i\Phi }}}
\end{array}} \right)\nonumber\\
&=& 2{S^2}\rho \sin 2\left[ \Theta  - \theta  - \left( {M - \mu } \right)t\right]\nonumber
\end{eqnarray}
\[{J_z} \sim \left( {\begin{array}{*{20}{c}}
{{\zeta ^ * }{e^{ - i\Phi }}}&{ - \zeta {e^{i\Phi }}}
\end{array}} \right)\left( {\begin{array}{*{20}{c}}
0&1\\
1&0
\end{array}} \right)\left( {\begin{array}{*{20}{c}}
{{\zeta ^ * }{e^{ - i\Phi }}}\\
{\zeta {e^{i\Phi }}}
\end{array}} \right) = 0\]
\[{J_t} \sim \left( {\begin{array}{*{20}{c}}
{{\zeta ^ * }{e^{ - i\Phi }}}&{ - \zeta {e^{i\Phi }}}
\end{array}} \right)\left( {\begin{array}{*{20}{c}}
0&{ - 1}\\
1&0
\end{array}} \right)\left( {\begin{array}{*{20}{c}}
{{\zeta ^ * }{e^{ - i\Phi }}}\\
{\zeta {e^{i\Phi }}}
\end{array}} \right) =  - 2{S^2}\]
If a Cochlean current with a purely azimuthal space component is to be constructed as a conformal multiple of the Riemann current, and if it is also to be constant in time, then it follows that 
\[\mu  = M \wedge \Theta  - \theta  = \left( {2N + 1} \right)\pi /4 ,\quad N \in {\cal Z}, \]
from which there also follow the respective Riemann and Cochlean pseudoequality relationships 
\[{J_\phi } = \varepsilon \rho {J_t} \Leftrightarrow {j_\phi } = \varepsilon \rho {j_t};\;\varepsilon  := {\left(  -  \right)^{N + 1}}.\]
It may be noted that the conformal factor that links these two currents involves the exponential decay factors in $\rho$ and $z$ that characterise the Cochlean spinor modulus $H$.

\section{Investigating the Riemann Tensor Connection}\label{Ch3}

\subsection{Determining the Connection}

The analysis at the spinor level fixed not only the general form of the Riemann spinor connection, it also established the forms of the IvdW; it is the covariant constancy of the latter that will now be used to determine the tensor connection using the fields $A$ and $U$ that appear in the spinor connection. Covariant constancy of the IvdW is expressed by
\[\begin{array}{l}
\quad \;{D_\mu }{\Sigma _{\nu A\dot B}} = 0\\
 \Leftrightarrow {\left( {\ln \left( {\Omega {S^2}} \right)} \right)_\mu }{\Sigma _{\nu A\dot B}} + \Omega {S^2}{{\hat \sigma '}_{\nu A\dot B\mu }} = \Gamma _{ \bullet \nu \mu }^\alpha {\Sigma _{\alpha A\dot B}} + \Gamma _{\mu A}^{\quad C}{\Sigma _{\nu C\dot B}} + \Gamma _{\mu \dot B}^{\quad \dot D}{\Sigma _{\nu A\dot D}}\\
 \Leftrightarrow {\omega _\mu }{\Sigma _{\nu A\dot B}} + \Omega {S^2}{{\hat \sigma '}_{\nu A\dot B\mu }} = \Gamma _{ \bullet \nu \mu }^\alpha {\Sigma _{\alpha A\dot B}} + \Gamma _{\mu (A}^{\quad C)}{\Sigma _{\nu C\dot B}} + \Gamma _{\mu (\dot B}^{\quad \dot D)}{\Sigma _{\nu A\dot D}}
\end{array}\]
when contracting with $\Sigma _\rho ^{\dot BA}$ gives
\begin{equation}\label{(3.1.2)}{\Gamma _{\rho \nu \mu }} = {\omega _\mu }{G_{\nu \rho }} + {\Omega ^2}{\hat \sigma '_{\nu A\dot B\mu }}\hat \sigma _\rho ^{\dot BA} - 2{\mathop{\rm Re}\nolimits} \left\{ {\Gamma _{\mu (A}^{\quad B)}\Sigma _{B\rho \nu }^{ \bullet A}} \right\}\end{equation}
It should be noted that the prime and extra lower spacetime index here indicate a standard coordinate derivative.
The simplification of (\ref{(3.1.2)}) involves lengthy manipulations that will not be given here, but lead to the result 
\begin{equation}\label{(3.1.3)}{\Gamma _{\rho \nu \mu }} = {\omega _\mu }{G_{\rho \nu }} + 2{\Omega ^2}\left\{ \begin{array}{l}
\rho {\mathop{\rm Im}\nolimits} {U_\mu }{\delta _{\rho \nu 21}} + \varepsilon {\mathop{\rm Im}\nolimits} {U_\mu }{\delta _{\rho \nu 41}}\\
 + {\mathop{\rm Re}\nolimits} {U_\mu }{\delta _{\rho \nu 34}} + \rho \varepsilon {\mathop{\rm Re}\nolimits} {U_\mu }{\delta _{\rho \nu 32}}\\
 + e\varepsilon {A_\mu }{\delta _{\rho \nu 31}} - \frac{\rho }{2}{\delta _{\mu 1}}{\delta _{\nu 2}}{\delta _{\rho 2}}
\end{array} \right\};\;{\delta _{\rho \nu ab}} := {\delta _{\rho a}}{\delta _{\nu b}} - {\delta _{\rho b}}{\delta _{\nu a}},\end{equation}
of which the torsion is
\[\begin{array}{l}
{T_{\rho \nu \mu }} := {\Gamma _{\rho \nu \mu }} - {\Gamma _{\rho \mu \nu }} = {\omega _\mu }{G_{\rho \nu }} - {\omega _\nu }{G_{\rho \mu }}\\
 + 2{\Omega ^2}\left\{ \begin{array}{l}
\rho {\mathop{\rm Im}\nolimits} {U_\mu }{\delta _{\rho \nu 21}} + \varepsilon {\mathop{\rm Im}\nolimits} {U_\mu }{\delta _{\rho \nu 41}}\\
 + {\mathop{\rm Re}\nolimits} {U_\mu }{\delta _{\rho \nu 34}} + \rho \varepsilon {\mathop{\rm Re}\nolimits} {U_\mu }{\delta _{\rho \nu 32}}\\
 + e\varepsilon {A_\mu }{\delta _{\rho \nu 31}} - \frac{\rho }{2}{\delta _{\mu 1}}{\delta _{\nu 2}}{\delta _{\rho 2}}
\end{array} \right\} - 2{\Omega ^2}\left\{ \begin{array}{l}
\rho {\mathop{\rm Im}\nolimits} {U_\nu }{\delta _{\rho \mu 21}} + \varepsilon {\mathop{\rm Im}\nolimits} {U_\nu }{\delta _{\rho \mu 41}}\\
 + {\mathop{\rm Re}\nolimits} {U_\nu }{\delta _{\rho \mu 34}} + \rho \varepsilon {\mathop{\rm Re}\nolimits} {U_\nu }{\delta _{\rho \mu 32}}\\
 + e\varepsilon {A_\nu }{\delta _{\rho \mu 31}} - \frac{\rho }{2}{\delta _{\nu 1}}{\delta _{\mu 2}}{\delta _{\rho 2}}
\end{array} \right\}
\end{array}.\]

If $U = A = 0$ and $\Omega = 1$, the Riemann connection and torsion reduce to the Cochlean forms of Paper 1,
\[{\breve \gamma } _{\rho \nu \mu } =  - \rho {\delta _{\rho 2}}{\delta _{\nu 2}}{\delta _{\mu 1}},\quad {\breve t} _{\rho \nu \mu } = \rho {\delta _{\rho 2}}{\delta _{\nu \mu 12}}.\]

\subsection{Riemann Torsion and Electromagnetism}\label{subsec3.2}

It is natural to propose that the Riemann definition of the electromagnetic field in terms of the putative Riemann potential $A_\mu$, have the standard form, so that 
\begin{equation}\label{(3.2.1)}{F_{\mu \nu }} := {D_\mu }{A_\nu } - {D_\nu }{A_\mu } = {d_\mu }{A_\nu } - {d_\nu }{A_\mu } + T_{ \bullet \mu \nu }^\sigma {A_\sigma }.\end{equation}
In simplifying the final term here the conformality of $A_\sigma$ with $a_\sigma$ guarantees that only components with $\sigma = 2, 4$ need be considered, when the contraction requires 
\[\label{(3.2.2)}T_{ \bullet \mu \nu }^2{A_2} = {A_2}{\omega _\nu }{\delta _{2\mu }} - {A_2}{\omega _\mu }{\delta _{2\nu }} - \frac{{{A_2}}}{\rho }\left[ \begin{array}{c}
2\left( {{\mathop{\rm Im}\nolimits} {U_\nu }{\delta _{\mu 1}} - {\mathop{\rm Im}\nolimits} {U_\mu }{\delta _{\nu 1}}} \right) + {\delta _{\mu \nu 12}}\\
 + 2\varepsilon \left( {{\mathop{\rm Re}\nolimits} {U_\mu }{\delta _{\nu 3}} - {\mathop{\rm Re}\nolimits} {U_\nu }{\delta _{\mu 3}}} \right)
\end{array} \right],\]
\[T_{ \bullet \mu \nu }^4{A_4} = {A_4}{\omega _\nu }{\delta _{4\mu }} - {A_4}{\omega _\mu }{\delta _{4\nu }} + 2{A_4}\left[ \begin{array}{c}
\left( {{\mathop{\rm Re}\nolimits} {U_\mu }{\delta _{\nu 3}} - {\mathop{\rm Re}\nolimits} {U_\nu }{\delta _{\mu 3}}} \right)\\
 + \varepsilon \left( {{\mathop{\rm Im}\nolimits} {U_\nu }{\delta _{\mu 1}} - {\mathop{\rm Im}\nolimits} {U_\mu }{\delta _{\nu 1}}} \right)
\end{array} \right].\]
Defining
\begin{equation}\label{(3.2.4)}{A_ + } := \frac{{{A_2}}}{\rho } + {\left(  -  \right)^{N + 1}}{A_4} = \frac{{{A_2}}}{\rho } - \varepsilon {A_4},\end{equation}
the six Riemann electromagnetic field components of (\ref{(3.2.1)}) then become
\begin{equation}\label{(3.2.5)}{F_{12}} = {d_1}{A_2} + T_{ \bullet 12}^\sigma {A_\sigma } = \rho \left[ {{d_1}\left( {\frac{{{A_2}}}{\rho }} \right) - {\omega _1}\left( {\frac{{{A_2}}}{\rho }} \right)} \right] - 2{A_ + }{\mathop{\rm Im}\nolimits} {U_2},\end{equation}
\begin{equation}\label{(3.2.6)}{F_{13}} = T_{ \bullet 13}^\sigma {A_\sigma } =  - 2{A_ + }\left( {{\mathop{\rm Im}\nolimits} {U_3} + \varepsilon {\mathop{\rm Re}\nolimits} {U_1}} \right),\end{equation}
\begin{equation}\label{(3.2.7)}{F_{14}} = {d_1}{A_4} + T_{ \bullet 14}^\sigma {A_\sigma } = {d_1}{A_4} - {\omega _1}{A_4} - 2{A_ + }{\mathop{\rm Im}\nolimits} {U_4},\end{equation}
\begin{equation}\label{(3.2.8)}{F_{23}} =  - {d_3}{A_2} + T_{ \bullet 23}^\sigma {A_\sigma } = \rho \left[ {{d_3}\left( {\frac{{{A_2}}}{\rho }} \right) - {\omega _3}\left( {\frac{{{A_2}}}{\rho }} \right)} \right] - 2\varepsilon {A_ + }{\mathop{\rm Re}\nolimits} {U_2},\end{equation}
\begin{equation}\label{(3.2.9)}{F_{24}} = 0,\end{equation}
\begin{equation}\label{(3.2.10)}{F_{34}} = {d_3}{A_4} + T_{ \bullet 34}^\sigma {A_\sigma } = {d_3}{A_4} - {\omega _3}{A_4} + 2\varepsilon {A_ + }{\mathop{\rm Re}\nolimits} {U_4}.\end{equation}

Now, if the Cochlean $a_\mu$ is to be conformal with the Riemann field $A_\mu$, this must also display the pseudoequality property that $A_2 = \varepsilon\rho {A_ 4}$, so that the composite field ${A_ + }$ defined in (\ref{(3.2.4)}) is identically zero and therefore vanishes from (\ref{(3.2.5)}) – (\ref{(3.2.10)}). The Riemann electromagnetic field itself can then be made to vanish if the choice is made that, with a notional dimensionless charge factor, $q$,
\begin{equation}\label{(3.2.11)}{A_4} = Mq\Omega \quad \left\langle { \Rightarrow {A_2} = \varepsilon \rho Mq\Omega } \right\rangle. \end{equation}
This condition will be assumed true from now on, even though for convenience in tracking the appearance of the $A_\mu$ field the substitutions implied here will not be made immediately.
Under these circumstances it follows that the electromagnetic field that is essential to a Poincaré account of the interactions of charged particles arises from a spacetime curvature and torsion that a Poincaré observer is unable to observe.

\section{Constructing the Energy-Momentum Tensor}\label{Ch4}

Essential properties of an energy-momentum density tensor in a Poincaré spacetime are that it be of second rank, symmetric and conserved. Whilst its construction, partly from the spinor current, ensured the tensor proposed in the Cochlean spacetime of Paper 1 automatically had these properties, it further had several individual elements that vanished, together with its trace; these properties are used to guide construction of its counterpart in the semi-Riemannian spacetime of the present paper. If torsion is present in such a spacetime then there are several ways of constructing a symmetric second-rank tensor that, as the torsion vanishes, reproduce the Einstein tensor that is effectively the energy-momentum density tensor of General Relativity. The required tensor will be constructed here by symmetrising and carrying out the standard trace subtraction on a Ricci tensor that is defined in the usual way. 

\subsection{Properties of the Ricci Tensor}

With the Riemann tensor connection given in (\ref{(3.1.3)}) the corresponding Ricci tensor, defined as
\[\label{(4.1.1)}{R_{\mu \nu }} = \Gamma _{ \bullet \mu \nu \beta }^{\beta  \bullet  \bullet } - \Gamma _{ \bullet \mu \beta \nu }^{\beta  \bullet  \bullet } + \Gamma _{ \bullet \mu \nu }^{\rho  \bullet  \bullet }\Gamma _{ \bullet \rho \beta }^{\beta  \bullet  \bullet } - \Gamma _{ \bullet \mu \beta }^{\rho  \bullet  \bullet }\Gamma _{ \bullet \rho \nu }^{\beta  \bullet  \bullet },\]
can be shown to be
\begin{equation}\label{(4.1.2)}\begin{array}{c}
{R_{\mu \nu }} =  - 4e{A_\nu }\left( {{\delta _{\mu 1}}{\mathop{\rm Re}\nolimits} {U_ + } + \varepsilon {\delta _{\mu 3}}{\mathop{\rm Im}\nolimits} {U_ + }} \right) - 2\varepsilon e\left( {{A_{\nu 3}}{\delta _{\mu 1}} - {A_{\nu 1}}{\delta _{\mu 3}}} \right)\\
 + \frac{{2{\delta _{\nu 1}}}}{{{\rho ^2}}}\left( {{\mathop{\rm Im}\nolimits} {U_2}{\delta _{\mu 1}} - \varepsilon {\mathop{\rm Re}\nolimits} {U_2}{\delta _{\mu 3}}} \right) - 2\varepsilon \left( {{\delta _{\mu 3}}{\mathop{\rm Re}\nolimits} {U_{ + \nu }} - \varepsilon {\delta _{\mu 1}}{\mathop{\rm Im}\nolimits} {U_{ + \nu }}} \right)\\
 - 2\left( {{\mathop{\rm Im}\nolimits} {U_{1\nu }} - \varepsilon {\mathop{\rm Re}\nolimits} {U_{3\nu }}} \right){\delta _{\mu 24}} + 2\left( {{\mathop{\rm Im}\nolimits} {U_{\nu 1}} - \varepsilon {\mathop{\rm Re}\nolimits} {U_{\nu 3}}} \right){\delta _{\mu 24}}\\
 + 4\varepsilon e{A_\nu }\left( {{\mathop{\rm Im}\nolimits} {U_3} + \varepsilon {\mathop{\rm Re}\nolimits} {U_1}} \right){\delta _{\mu 24}},
\end{array}\end{equation}
where
\[{U_ + } := \frac{{{U_2}}}{\rho } + {\left(  -  \right)^{N+1}}{U_4} = \frac{{{U_2}}}{\rho } - \varepsilon {U_4},\quad {\delta _{\mu 24}} := \rho {\delta _{\mu 2}} + \varepsilon {\delta _{\mu 4}},\]
and the identification of $A_{\mu}$ in (\ref{(3.2.11)}) remains to be implemented.
As just noted, the generalised Einstein tensor will be chosen to be defined from the symmetrised form of $R_{\mu\nu}$ with the usual trace subtraction so, using the standard notation for it,
\begin{equation}\label{(4.1.4)}{E_{\mu \nu }} := {R_{\left( {\mu \nu } \right)}} - \frac{1}{2}R{G_{\mu \nu }};\;R := {G^{\mu \nu }}{R_{\mu \nu }};\;E := {G^{\mu \nu }}{E_{\mu \nu }} =  - R.\end{equation}
In order to evaluate the trace, consider the diagonal elements of (\ref{(4.1.2)}), which are
\begin{equation}\label{(4.1.5)}{R_{11}} = \frac{2}{{{\rho ^2}}}{\mathop{\rm Im}\nolimits} {U_2} + 2{d_\rho }{\mathop{\rm Im}\nolimits} {U_ + },\end{equation}
\[{R_{22}} = 2\rho \left( {{\mathop{\rm Im}\nolimits} {U_{21}} - \varepsilon {\mathop{\rm Re}\nolimits} {U_{23}}} \right) + 4\varepsilon e\rho {A_2}\left( {{\mathop{\rm Im}\nolimits} {U_3} + \varepsilon {\mathop{\rm Re}\nolimits} {U_1}} \right),\]
\begin{equation}\label{(4.1.7)}{R_{33}} =  - 2\varepsilon {d_z}{\mathop{\rm Re}\nolimits} {U_ + },\end{equation}
\[{R_{44}} = 2\varepsilon \left( {{\mathop{\rm Im}\nolimits} {U_{41}} - \varepsilon {\mathop{\rm Re}\nolimits} {U_{43}}} \right) + 4e{A_4}\left( {{\mathop{\rm Im}\nolimits} {U_3} + \varepsilon {\mathop{\rm Re}\nolimits} {U_1}} \right).\]
Using these forms, the trace is given by
\begin{eqnarray}
{\Omega ^2}R & = & {R_{44}} - {R_{33}} - \frac{1}{{{\rho ^2}}}{R_{22}} - {R_{11}}\nonumber\\
 &=& \left[ \begin{array}{c}
2\varepsilon \left( {{\mathop{\rm Im}\nolimits} {U_{41}} - \varepsilon {\mathop{\rm Re}\nolimits} {U_{43}}} \right) + 2\varepsilon {d_z}{\mathop{\rm Re}\nolimits} {U_ + }\\
 - \frac{2}{\rho }\left( {{\mathop{\rm Im}\nolimits} {U_{21}} - \varepsilon {\mathop{\rm Re}\nolimits} {U_{23}}} \right) - \frac{2}{{{\rho ^2}}}{\mathop{\rm Im}\nolimits} {U_2} - 2{d_\rho }{\mathop{\rm Im}\nolimits} {U_ + }
\end{array} \right]\nonumber\\
& =& 4\varepsilon {d_z}{\mathop{\rm Re}\nolimits} {U_ + } - 4{d_\rho }{\mathop{\rm Im}\nolimits} {U_ + } - \frac{4}{{{\rho ^2}}}{\mathop{\rm Im}\nolimits} {U_2} =  - 2\left( {{R_{33}} + {R_{11}}} \right).\nonumber\end{eqnarray}
The {\em ad hoc} Cochlean energy momentum density tensor not only has vanishing trace, but its $(3,3)$ and $(1,1)$ elements also vanish; the generalised conformality of Cochlean and Riemann forms of the energy tensor suggests that similar conditions be imposed on the Ricci tensor. Vanishing of $R_{33}$ requires from (\ref{(4.1.7)}) that the real part of $U_+$ be a function of $\rho$ only, while vanishing of $R_{11}$ requires from (\ref{(4.1.5)}) that ${\rm Im}U_2$ and ${\rm Im}U_4$ satisfy
\begin{equation}\label{(4.1.10)}\frac{1}{{{\rho ^2}}}{\mathop{\rm Im}\nolimits} {U_2} + {d_\rho }{\mathop{\rm Im}\nolimits} {U_ + } = \frac{{{d_\rho }{\mathop{\rm Im}\nolimits} {U_2}}}{\rho } - \varepsilon {d_\rho }{\mathop{\rm Im}\nolimits} {U_4} = 0,\end{equation}
a property that can be ensured by defining the arbitrary ${\rm Im}U$components in terms of an equally arbitrary real function $f$
\begin{equation}\label{(4.1.11)}{\mathop{\rm Im}\nolimits} {U_ + } := f\left( \rho  \right) \Rightarrow {\mathop{\rm Im}\nolimits} {U_2} =  - {\rho ^2}f'\left( \rho  \right) \wedge {\mathop{\rm Im}\nolimits} {U_4} =  - \varepsilon \left( {\rho f'\left( \rho  \right) + f\left( \rho  \right)} \right).\end{equation}
Noting also the conditions $t_{22} = \varepsilon\rho t_{24} = {\rho}^2 t_{44}$ satisfied by the Cochlean energy density tensor, conformality requires a traceless Ricci tensor to satisfy $R_{22} = \varepsilon\rho R_{(24)} = {\rho}^2 R_{44}$. From (\ref{(4.1.2)}) it follows that 
\[{R_{22}} - \varepsilon \rho {R_{\left( {24} \right)}} \propto {R_{\left( {24} \right)}} - \varepsilon \rho {R_{44}} \propto {d_z}{\mathop{\rm Re}\nolimits} {U_ + } - \varepsilon \left( {{d_\rho }{\mathop{\rm Im}\nolimits} {U_2} - \varepsilon \rho {d_\rho }{\mathop{\rm Im}\nolimits} {U_4}} \right),\]
when the vanishing of $R_{11}$ and $R_{33}$ from (\ref{(4.1.10)}) and (\ref{(4.1.7)}) respectively ensures the required properties of $R_{\mu\nu}$ for even values of $\mu$ and $\nu$. 
 
\subsection{Constraining the Einstein Tensor}

Noting the simple multiplicative constant in the classical Einstein condition, 
\begin{equation}\label{(4.2.1)}{E_{\mu \nu }} = 8\pi N{T_{\mu \nu }} = \frac{{8\pi }}{{M_P^2}}{T_{\mu \nu }},\end{equation}
where $M_P$ is the Planck mass, effectively identifies the Einstein tensor with the energy-momentum density tensor. In the absence of torsion, the Einstein tensor is automatically conserved, as required for the physical tensor. In the current formulation however, the generalised form (\ref{(4.1.4)}) must be explicitly constrained to be divergence-free,
\begin{equation}\label{(4.2.2)}{\partial _\nu } := {D_\mu }{G^{\mu \alpha }}{E_{\alpha \nu }} = {G^{\mu \alpha }}\left( {{d_\mu }{E_{\alpha \nu }} - \Gamma _{ \bullet \alpha \mu }^{\beta  \bullet  \bullet }{E_{\beta \nu }} - \Gamma _{ \bullet \nu \mu }^{\beta  \bullet  \bullet }{E_{\alpha \beta }}} \right) = 0.\end{equation}
There are four somewhat complex conditions here, distinguished by the value of $\nu$; combining those for the  $\nu$ values of 2 and 4 gives the simpler necessary condition 
\begin{equation}\label{(4.2.3)}0 = \varepsilon \rho {\kern 1pt} {\partial _4} - {\partial _2} = \frac{{e{A_4}}}{{{\Omega ^2}}}\left[ {{d_z}\ln \left| {{A_4}} \right| + 2\frac{\varepsilon }{\rho }\left( {{\mathop{\rm Re}\nolimits} {U_ + } - \varepsilon \rho {\omega _z}} \right)} \right] = \varepsilon \frac{{e{A_4}}}{{{\Omega ^2}\rho }}\left( {2{\mathop{\rm Re}\nolimits} {U_ + } - \varepsilon \rho {\omega _z}} \right),\end{equation}
where, according to (\ref{(3.2.11)}), $A_4$ has been taken to be a multiple of $\Omega$. 
Recalling that for the vanishing of $R_{33}$ in (\ref{(4.1.7)}) ${\rm Re}U_+$ must be a function of $\rho$ only, $\omega_z$ must be zero here to avoid exponential growth or decay of $\Omega$ for asymptotic $z$, and this requires further that ${\rm Re}U_+$ vanish. The vanishing of $\omega_z$ makes the metric conformal factor a function of $\rho$ only, and this leads to considerable simplifications in the components (\ref{(4.2.2)}).
Taking in particular $\nu = 1$ now gives
\begin{equation}\label{(4.2.4)}e{A_4} + {\mathop{\rm Im}\nolimits} {U_3} = 0 \Leftrightarrow {\mathop{\rm Im}\nolimits} {U_3} =  - e{A_4},\end{equation}
whilst $\nu = 3$ gives
\begin{equation}\label{(4.2.5)}\varepsilon {\mathop{\rm Im}\nolimits} {U_1} = 2{\mathop{\rm Re}\nolimits} {U_3}.\end{equation}
When these constraints are used to simplify either ${\partial _2 }$  or ${\partial _4 }$, it follows that  
\begin{equation}\label{(4.2.6)}0 =  - 2\varepsilon {A_4}{\mathop{\rm Re}\nolimits} {U_4} + 2\rho \left( {2{f_1} + {\omega _\rho }} \right)\left( {{\mathop{\rm Re}\nolimits} {U_{3\rho }} - {\mathop{\rm Re}\nolimits} {U_{1z}}} \right) + \rho \left( \begin{array}{c}
2{\mathop{\rm Re}\nolimits} {U_{3zz}}\\
 + {\mathop{\rm Re}\nolimits} {U_{1\rho z}} - {\mathop{\rm Re}\nolimits} {U_{3\rho \rho }}
\end{array} \right).\end{equation}

Moving on from the conservation conditions (\ref{(4.2.3)}) – (\ref{(4.2.6)}), further constraints arise on $E_{\mu\nu}$ from the nature of the state described by the physical spinor, namely that it be that of a spin-1/2 fermion at rest at the origin of the space coordinate system. These constraints concern only the off-diagonal elements of $E_{\mu\nu}$, which from (\ref{(4.1.2)}) are
\begin{equation}\label{(4.2.7)}{E_{12}} = \varepsilon \left( {\rho \left( {{\mathop{\rm Re}\nolimits} {U_{3\rho }} - {\mathop{\rm Re}\nolimits} {U_{1z}}} \right) - e{A_4}\left( {2{\mathop{\rm Re}\nolimits} {U_ + } + \varepsilon \rho {\omega _z}} \right)} \right) = \varepsilon \rho {E_{14}},\end{equation}
\[{E_{13}} =  - \frac{\varepsilon }{\rho }{d_\rho }{\mathop{\rm Re}\nolimits} {U_ + },\]
\[{E_{23}} = e{A_4} + \rho \left( {{\mathop{\rm Im}\nolimits} {U_{3\rho }} - {\mathop{\rm Im}\nolimits} {U_{1z}} - e{A_4}\left( {2{f_1} - {\omega _\rho }} \right)} \right) = e{A_4} + \varepsilon \rho {E_{34}},\]
\[{E_{24}} = \varepsilon \rho {E_{44}}.\]

To begin with, the overall momentum of the state must vanish. This property is in fact guaranteed by the azimuthal uniformity of the spinor modulus and evenness in the $z$ variable, which cause the required cancellations in the integrals of the momentum density components $E_{41}$, $E_{42}$ and $E_{43}$. 

Considering next angular momentum, the associated Riemann integrand may be formed by taking the skew element
\begin{equation}\label{(4.2.11)}{M_{4\mu \nu }} := {E_{4\alpha }}{x_\beta }e_{ \bullet  \bullet \mu \nu }^{\alpha \beta  \bullet  \bullet },\quad \;{e_{\alpha \beta \gamma \delta }} = \sqrt { - G} {\varepsilon _{\alpha \beta \gamma \delta }} = \rho \,{\Omega ^4}{\rm{\varepsilon }_{\alpha \beta \gamma \delta }},\end{equation}
where $\rm{\varepsilon}$ here is the totally skew symbol with $\rm{\varepsilon}_{1234} = 1$. Taking $\mu$  and $\nu$  to be 3 and 4 to obtain the $z$ component requires $\alpha$  and $\beta$  to be $[1,2]$ or $[2,1]$, and to obtain a standard form based upon the azimuthal momentum density given by $E_{42}$ the term in $E_{41}$ must vanish. Taking this with (\ref{(4.2.3)}) and (\ref{(4.2.7)}), gives
${\mathop{\rm Re}\nolimits} {U_{3\rho }} - {\mathop{\rm Re}\nolimits} {U_{1z}} = 0,$ causing (\ref{(4.2.6)}) to simplify to
\[0 =  - \varepsilon {A_4}{\mathop{\rm Re}\nolimits} {U_4} + \rho {\mathop{\rm Re}\nolimits} {U_{3zz}} \Leftrightarrow {\mathop{\rm Re}\nolimits} {U_{3zz}} = \frac{{{A_4}}}{{{\rho ^2}}}{\mathop{\rm Re}\nolimits} {U_2}.\]

When all these constraints are taken together there emerges a Riemann energy density of form 
\[{E_{44}} = 2\varepsilon \left( {{\mathop{\rm Im}\nolimits} {U_{4\rho }} - \varepsilon {\mathop{\rm Re}\nolimits} {U_{4z}}} \right) + 4e{A_4}\left( {{\mathop{\rm Im}\nolimits} {U_3} + \varepsilon {\mathop{\rm Re}\nolimits} {U_1}} \right).\]
To ensure asymptotic $z$-convergence of the energy integral any purely $\rho$-dependent term here must vanish; this fixes the form of $f(\rho)$ in (\ref{(4.1.11)}) and leads finally to
\begin{equation}\label{(4.2.14)}{E_{44}} = 2\left( {2e\varepsilon {A_4}{\mathop{\rm Re}\nolimits} {U_1} - {\mathop{\rm Re}\nolimits} {U_{4z}}} \right).\end{equation}
Since ${\rm Re}U_1$ and  ${\rm Re}U_4$  are arbitrary functions it follows that, despite the appearance of the Planck mass in the Einstein condition (\ref{(4.2.1)}), a mass scale has not yet been fixed in the current formulation. 
The overall scale of $E_ {\mu \nu}$ is not however free, since it is reasonable to require that spin, being dimensionless, remain at $1/2$ in the Riemann formulation, and this involves an integral of $E_{42} = \varepsilon\rho E_{44}$ that will next be considered.

\section{Establishing a Mass Scale}\label{Ch5}

The spin of the Riemann state being considered in this paper is
\begin{equation}\label{(5.1.1)}{L_z} = \int\limits_{\scriptstyle all\atop
\scriptstyle space} {{\Omega ^8}{T_{42}}d\tau }  = \varepsilon \int\limits_{\scriptstyle all\atop
\scriptstyle space} {{\Omega ^8}\rho \,{T_{44}}d\tau }, \end{equation}
where a weighting factor of ${\Omega ^4}$ comes from the definition (\ref{(4.2.11)}) and a similar one is provided by the Jacobian in the volume element.
The Cochlean state of Paper 1, on the other hand, had a spin of 
\[{\ell _z} = \varepsilon \int\limits_{\scriptstyle all\atop
\scriptstyle space} {\rho \,{{\breve t} }_{44}}d\tau  = \frac{\varepsilon }{2}\frac{{{m_{calc}}}}{{\left| m \right|\left( {1 - \cos \delta } \right)}} = \frac{\varepsilon }{2},\]
where the third equality arises by making an obvious choice for the relationship between formulational mass, $m$, and calculated mass, $m_{calc}$.

Noting that the $T_{44}$ obtained from (\ref{(4.2.14)}) is an arbitrary function of $\rho$ and $z$, a simple way of ensuring that it give a spin of 1/2 in (\ref{(5.1.1)}) is to choose it to be
\[{T_{44}} = {\Omega ^{ - 8}}{\breve t}_{44},\]
which then gives what may be termed the ‘Riemann mass’ to be
\begin{equation}\label{(5.1.4)}{M_R} := \int\limits_{\scriptstyle all\atop
\scriptstyle space} {{\Omega ^4}{T_{44}}d\tau }  = \int\limits_{\scriptstyle all\atop
\scriptstyle space} {{\Omega ^{ - 4}}{{\breve t} }_{44}}d\tau . \end{equation}

Expressing the formulational Cochlean mass, $m$, and the mass dependence of $\Omega$ in terms of the Planck mass, $M_P$, and the Riemann mass, $M_R$, causes (\ref{(5.1.4)}) to become a self-consistency condition that can be used to determine $M_R$ in terms of $M_P$, thereby establishing a mass scale. The result of such a procedure will of course be critically dependent on the model used for $\Omega$, insight into which may be obtained from the following reflections.

Suppose that the Cochlean formulational mass were taken to be $M_P$, while $M_R$ is understood to be effectively a physical particle mass, essentially that measured by a Poincaré observer; a typical such mass, that of the electron, would give the ratio $M_R /M_P$ to be of the order of $10^{-23}$. The Cochlean energy density in (\ref{(5.1.4)}) involves decaying exponentials with exponents in $M_P \rho$ and $M_P z$ that, in the absence of gravitation, give an overall factor on the RHS of $M_P$. Obtaining a value for $M_R /M_P$ that is extremely small requires that $\Omega$ in (\ref{(5.1.4)}), a function of $\rho$ only, be extremely large for small $\rho$, whilst going to unity of course asymptotically. 
A model for $\Omega$ that potentially has such properties is
\begin{equation}\label{(5.1.5)}{\Omega _f} = \exp \left( {\frac{{{\Gamma ^2}}}{{64{M_f}\rho }}} \right),\end{equation}
where $M_f$ is the fermion mass and the real parameter $\Gamma^2$ that appears represents a gravitational constant yet to be determined; the 64 and the square of the dimensionless  $\Gamma$ are used here simply for later calculational convenience. In order to fix $\Gamma$, the fermion mass here will be replaced by $M_R$, and a natural self-consistency condition sought for which this mass is taken as that of the idealised state described by the current simple, Cochlean-based formulation. It should be noted that only a single, idealised particle can be described by this simple model based on the current Cochlean form; an approach that includes observed spin-1/2 fermions must then be based upon a generalisation of the Cochlean treatment that distinguishes such particles. 

Lest it be thought that the proposal (\ref{(5.1.5)}) is entirely {\em ad hoc}, it should be noted that the discussion of the $\Omega$-like $S^2$ that follows (\ref{(2.3.3)}) may be interpreted as suggesting an exponential form for $\Omega$ with an exponent that is a squared solution of the Cochlean spinor identity. A purely  $\rho$-dependent such solution has the form, with arbitrary constants $A$ and $B$, 
\[H = \frac{{A + B\exp \left( { - 2{M_f}\rho } \right)}}{{\sqrt {{M_f}\rho } }},\]
and taking $B$ to vanish to obtain (\ref{(5.1.5)}) is a convenient simplification for an initial approach, immediately yielding integrals that involve Bessel functions. Indeed, using (\ref{(5.1.4)}) and (\ref{(5.1.5)}) with the Cochlean energy density form from Paper 1 with $\left| m \right| = {M_P}$, there follows, with $r = {M_P}\rho \left( {1 - \cos \delta } \right) + {M_P}z\sin \delta $,
\begin{eqnarray}
\frac{{{M_R}}}{{{M_P}}} &=& 64M_P^3\sin \delta {\left( {1 - \cos \delta } \right)^3}\int\limits_{0,\infty } {\rho \exp \left( { - 4r - \frac{{{\Gamma ^2}}}{{16{M_R}\rho }}} \right)\,dzd\rho } \nonumber \\
 &= &\left( {1 - \cos \delta } \right)\int\limits_{0,\infty } {x\exp \left( { - x - \frac{{{{\hat \Gamma }^2}}}{{4x}}} \right)\,dx} ,\quad {{\hat \Gamma }^2} := \frac{{{M_P}\left( {1 - \cos \delta } \right)}}{{{M_R}}}{\Gamma ^2}\nonumber\\
& = &\left( {1 - \cos \delta } \right) \times \frac{1}{2}{{\hat \Gamma }^2}{K_2}\left( {\hat \Gamma } \right)\quad \mathop  \to \limits_{\Gamma  \to 0} \quad 1 - \cos \delta  \nonumber
\end{eqnarray}
where $K_2$ is a second-order modified Bessel function \cite {A&S}.
Defining now $M = \frac{{{M_R}}}{{1 - \cos \delta }}$, it follows that
\begin{equation}\label{(5.1.8)}\frac{M}{{{M_P}}} = \frac{1}{2}{\hat \Gamma ^2}{K_2}\left( {\hat \Gamma } \right),\quad {\hat \Gamma ^2} := \frac{{{M_P}}}{M}{\Gamma ^2}.\end{equation}
In order to set this up as a self-consistency condition a constant value is required for $\Gamma $. Such a value may be found by first noting the suggestive way electromagnetism, in the guise of the fine-structure constant $\alpha$, can be used to mediate between the Planck mass and, for example, the electron mass as it appears in  
\[{\left( {\frac{{{M_P}}}{{{M_e}}}} \right)^\alpha } \simeq 1.46,\quad {\left( {\frac{{{M_P}}}{{{M_e}}}} \right)^{\sqrt \alpha  }} \simeq 81.6.\]
Given the rapid decrease of the Bessel function as its argument increases from zero, (\ref{(5.1.8)}) requires that $\hat \Gamma$  be large, but not very large, perhaps close to 81.6. Indeed, noting this value, as well as bearing in mind the observations of Section \ref{subsec3.2} on the interdependence of gravitation and electromagnetism, one might take 
\[\hat \Gamma  = {\left( {\frac{{{M_P}}}{M}} \right)^{\sqrt \alpha  }},\]
when (\ref{(5.1.8)}) becomes
\[{\hat \Gamma ^{ - \frac{1}{{\sqrt \alpha  }}}} = \frac{1}{2}{\hat \Gamma ^2}{K_2}\left( {\hat \Gamma } \right),\]
of which the solution to six figures is 51.6641, giving
\begin{equation}\label{(5.1.12)}\Gamma  = \sqrt {\frac{M}{{{M_P}}}} \hat \Gamma  = {\hat \Gamma ^{1 - 1/\sqrt \alpha  }} \simeq 4.55190 \times {10^{ - 19}}.\end{equation}
It may be noted that the ${\Gamma}^2/64$  that appears in the metric conformal factor (\ref{(5.1.5)}) is roughly of the order of the ratio of the gravitational to the electrostatic force between typical pairs of elementary particles \cite{A&K}, again consonant with connection between gravity and electromagnetism.
Whatever interpretation is put upon $\Gamma $ however, if it has the value (\ref{(5.1.12)}) there then follows that
\[\frac{{{M_R}}}{{{M_P}}} \simeq \left( {1 - \cos \delta } \right) \times 8.81057 \times {10^{ - 21}},\]
and a choice of the formulational Cochlean $\delta $ close to $0.1$ may be used to make $M_R$ equal to the electron mass, thereby providing a satisfactory notional mass scale and effectively completing the purely formulational work of the current paper.

\section{Summary and Outlook}\label{Ch6}

This paper has shown that an earlier formulation of the description of charged spin-1/2 fermions in flat spacetimes with torsion can be developed into a pseudo-Riemannian formulation that uses a conformally Minkowskian metric. In the curved spacetime the electromagnetic field of the particle has been shown to vanish when the electrostatic potential is taken to be a constant multiple of the metric conformal factor. Gravitation and electromagnetism therefore appear to be closely connected, and this idea is reinforced in Section \ref{Ch5}, where the Planck mass and the fine-structure constant together provide a realistic mass scale, albeit for an idealised single-particle state.

Following on from these reflections, the electromagnetic structures of real fermions as they are observed in the underlying flat spacetime must be understood as being geometrically generated, partly by torsion and partly by the conformal factor of the Riemann metric, so that details of particle properties must be sought at a putative Poincaré level that somehow combines the pure Cochlean and Riemann treatments of these first two papers; it is hoped to present such a Poincaré analysis in a third paper that will complete the current series.

\end{document}